\documentclass[aps,prb,twocolumn,showpacs,amssymb,amsmath,superscriptaddress,floatfix]{revtex4}
\usepackage{graphics,epsfig,subfigure,graphicx}

\begin{document}

\title{Critical currents for vortex defect motion in superconducting arrays}
\author{Jong Soo Lim}
\affiliation{Department of Physics, Seoul National University,
Seoul 151-747, Korea}
\author{M.Y. Choi}
\affiliation{Department of Physics, Seoul National University,
Seoul 151-747, Korea}
\affiliation{Korea Institute for Advanced Study, Seoul 130-722, Korea.}
\author{Beom Jun Kim}
\affiliation{Department of Molecular Science and Technology, Ajou University,
Suwon 442-749, Korea}
\author{J. Choi}
\affiliation{Department of Physics, Keimyung University, Daegu
704-701, Korea}

\begin{abstract}
We study numerically the motion of vortices in two-dimensional
arrays of resistively shunted Josephson junctions. An extra vortex
is created in the ground states by introducing novel boundary conditions and 
made mobile by applying external currents. We then measure critical currents 
and the corresponding pinning energy barriers to vortex motion, 
which in the unfrustrated case agree well with previous theoretical and 
experimental findings. In the fully frustrated case our results also give 
good agreement with experimental ones, in sharp contrast with the existing
theoretical prediction. A physical explanation is provided in
relation with the vortex motion observed in simulations.
\end{abstract}


\maketitle

In equilibrium the two-dimensional Josephson junction array (JJA) 
is well known to exhibit the Berezinskii-Kosterlitz-Thouless transition,\cite{review} 
which is driven by unbinding of vortex-antivortex pairs. 
Unbound vortices, while in motion, dissipate energy and thus make the
system resistive. Without an external magnetic field this happens
above the critical temperature. It is possible, however,
to induce an unpaired vortex (defect) at low temperatures,
e.g., by applying weak magnetic fields. Such a field-induced vortex 
defect sits on a plaquette where  energy takes its local minimum; 
this is separated by the potential barrier set by the underlying lattice. 
The defect may be driven into motion by external currents, as it is exerted 
by the force
${\bf F} = (h/2e){\bf J}\times\hat{\bf z}$
with ${\bf{J}}$ being the external current density on the $xy$ plane.
When the driving force is sufficiently strong, the defect can  overcome the barrier 
to move onto the adjacent plaquette and accordingly give rise to non-vanishing voltage.
Theoretical calculation of the barrier height is based on the
iteration method, static in nature.\cite{Lobb,Mina} 
On the other hand, in experiment, the barrier height is determined dynamically, 
by measuring critical depinning currents.\cite{Rzchowski} 

In the absence of background frustration $f=0$, where $f$ measures the number of 
flux quanta per plaquette, this barrier has been estimated as $E_B(f{=}0)\approx 0.199$ 
in units of the Josephson coupling energy $E_J$.\cite{Lobb,Mina} The critical
current in this case has been obtained from the onset of instability of 
the phase configuration and found to be $I_c(f{=}0)\approx 0.105$ in units of the 
single-junction critical current $i_c$, which agrees well with experiment.\cite{Rzchowski} 
For high magnetic fields, say, $f=1/2$,
however, the numerically estimated value does not agree with the experimental one: 
$E_B(1/2)/E_B(0) \approx  6$ for the former whereas $E_B(1/2)/E_B(0) \approx 1.3$ 
for the latter. 
This discrepancy has remained unexplained since then.

This work is to bridge the gap between theory and experiment:
We perform extensive dynamic simulations of the resistively-shunted
junction (RSJ) model, which conveniently describes real dynamics of JJAs.
Unlike existing studies, which were limited mostly to the defect-free 
case because of the difficulty to create a single defect in the presence of 
external currents, we introduce novel boundary conditions for a single {\em extra}
vortex and study its motion in the ground state of the $f=1/2$ system as well as
the $f=0$ one. 
We then measure the critical current $I_c$ and
the corresponding energy barrier $E_B$ to vortex defect motion; this is,
to the best of our knowledge, the first direct numerical {\em measurement}
of the barrier height. 
It is confirmed that our results for $f=0$ agree well with
existing numerical and experimental values;\cite{Rzchowski,Benz}
also revealed in the fully-frustrated ($f=1/2$) JJA is excellent agreement 
of our results with experimental ones unexplained so far. 
Via direct observation of actual vortex motion, we suggest that the previous
overestimation of the energy barrier can be attributed to the
unrealistic configuration of vortices.

We begin with an $L\times L$ square array of RSJs between superconducting grains, 
with shunt resistance $R$. 
The $i$th grain, located at position ${\bf r}_i$, is described by its phase $\{\phi_i\}$.
%
%
In order to compute the energy barrier, it is necessary to create just one extra
vortex in the system. Both periodic boundary conditions (PBC)
and fluctuating twist boundary conditions (FTBC),\cite{Kim} used most frequently, 
allow $f = n/L$ with an integer $n$ and thus produce at the least $L$ extra vortices.
This problem can be remedied by using the diagonally antiperiodic boundary conditions (DAPBC) 
that introduce $2\pi$ rotations of the phase variables around the whole system and thus 
produce a single extra vortex.\cite{Kawamura} 
However, the straightforward application of DAPBC to dynamic simulations fails in the presence 
of external currents since the global current conservation cannot be fulfilled.\cite{Kim} 
We thus modify FTBC in such a way that the system contains one additional vortex. 
In FTBC, the gauge-invariant phase difference is decomposed into two parts: 
One is periodic across the system and thus does not carry voltage whereas the other, 
called the fluctuating twist variable, is directly related to the voltage drop across the system. 
On the former part the boundary conditions of the DAPBC type, detailed in Ref.~\onlinecite{bc}, 
are then imposed, relating only the boundary phase variables on the same columns and rows; 
this makes the application of the FTBC straightforward.  

Equations of motion for the system (without capacitive and inductive couplings) then read, 
after scaling the time in units of $\hbar /2ei_c R$,
\begin{equation}
{\sum_j}' \left[ \frac{d{\widetilde{\phi}}_{ij}}{dt} +
\sin({\widetilde{\phi}}_{ij} - {\bf r}_{ij}\cdot{\bf \Delta}) \right] = 0,
\label{eq:phi}
\end{equation}
where the primed summation runs over the nearest neighbors of grain $i$,
${\bf r}_{ij} \equiv {\bf r}_i - {\bf r}_j$ is a unit vector
with the lattice constant set to unity, 
and ${\widetilde{\phi}}_{ij} \equiv \phi_i - \phi_j - A_{ij}$, 
together with the voltage-carrying part, describes
the gauge-invariant phase difference between sites $i$ and $j$.
The bond angle $A_{ij}$, given by the line integral of the vector potential, 
takes in the Landau gauge the values
\begin{equation*}
A_{ij} =
\begin{cases}
0       & \textrm{for ${\bf r}_j = {\bf r}_i + {\bf \hat{x}}$}  \\
2\pi f x_i & \textrm{for ${\bf r}_j = {\bf r}_i + {\bf \hat{y}}$}
\end{cases}
\end{equation*}
with ${\bf r}_i$ denoting the position of site $i$.
The time evolution of the twist variable ${\bf \Delta} \equiv (\Delta_x, \Delta_y)$ 
is governed by\cite{Kim}
\begin{align}
& \frac{d\Delta_x}{dt}
 = \frac{1}{L^2} \sum_{{\langle ij \rangle}_x}\sin({\widetilde{\phi}}_{ij} - \Delta_x) - I_{dc}\nonumber \\
& \frac{d\Delta_y}{dt}
  = \frac{1}{L^2} \sum_{{\langle ij \rangle}_y}\sin({\widetilde{\phi}}_{ij} - \Delta_y),
\label{eq:D}
\end{align}
where $\sum_{\left\langle ij \right\rangle_a}$ denotes the
summation over all nearest neighboring pairs in the $a \,(= x, y)$
direction and the dc current $I_{dc}$ (measured in units of $i_c$) is injected along the $x$
direction.

\begin{figure}
\centering
\mbox{%
\subfigure{%
\includegraphics[width=2.5cm]{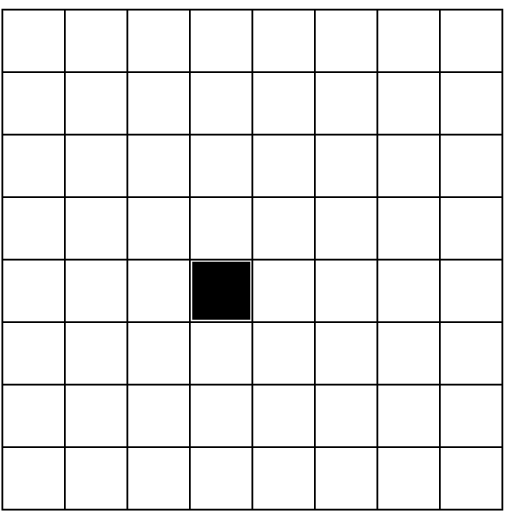}
}\quad
\subfigure{%
\includegraphics[width=2.5cm]{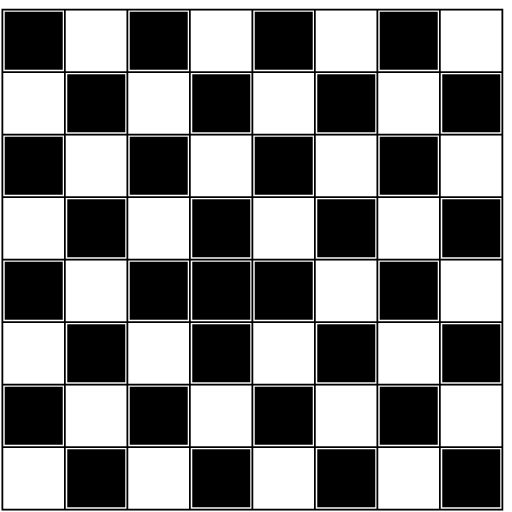}
}} 
\caption[Zero-temperature configuration] 
{Zero-temperature vortex configurations under the boundary conditions in Ref.~\onlinecite{bc} 
for $f=0$ (left) and $f=1/2$ (right).  Filled squares denote the plaquettes occupied by vortices.}
\label{fig:ground}
\end{figure}

We use the second-order Euler algorithm to integrate Eqs.~(\ref{eq:phi}) and (\ref{eq:D}) 
with the time step of size $\Delta t = 0.05$. 
As initial conditions, we use the well-known ground-state phase configurations 
obtained with the conventional PBC:
the ferromagnetic configuration $\phi_i = 0$ for $f=0$ and the checker-board configuration 
${\widetilde{\phi}}_{ij}=\pi/4$ for $f=1/2$. 
After appropriate equilibration, the zero-temperature vortex configurations in
Fig.~\ref{fig:ground} are obtained, which manifestly show the location of the 
extra vortex and justify the boundary conditions employed. 
We then apply the external dc current $I_{dc}$ and measure the voltage drop (in units of $i_c R$)
\begin{equation}
\langle V \rangle = -\frac{\hbar L}{2e} \left\langle\frac{d\Delta_x}{dt}\right\rangle,
\label{eq:V}
\end{equation}
with $\langle \cdots \rangle$ denoting the time average
and the energy
\begin{equation}
E  = - \sum_{\langle ij \rangle} \cos({\widetilde{\phi}}_{ij} - {\bf r}_{ij}\cdot{\bf \Delta}).
\label{eq:E}
\end{equation}
in units of the Josephson coupling strength $E_J \equiv \hbar i_c /2e$. 
Typically, data for physical quantities have been averaged over $5\times 10^5$
time steps after equilibration over $3\times 10^4$ time steps for
the JJA with size up to $L=120$.

\begin{figure}
\centering 
\epsfig{width=0.4\textwidth,file=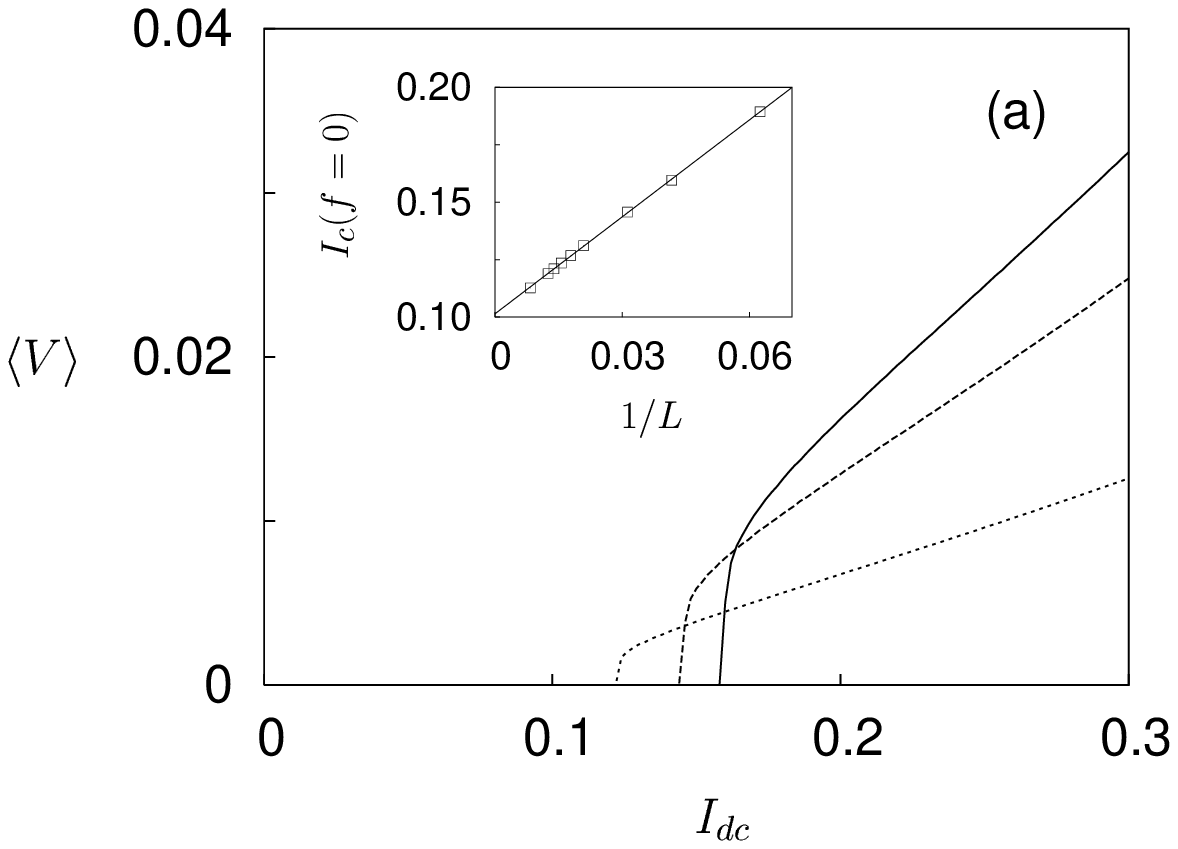}
\vspace{0.3cm} 
\epsfig{width=0.4\textwidth,file=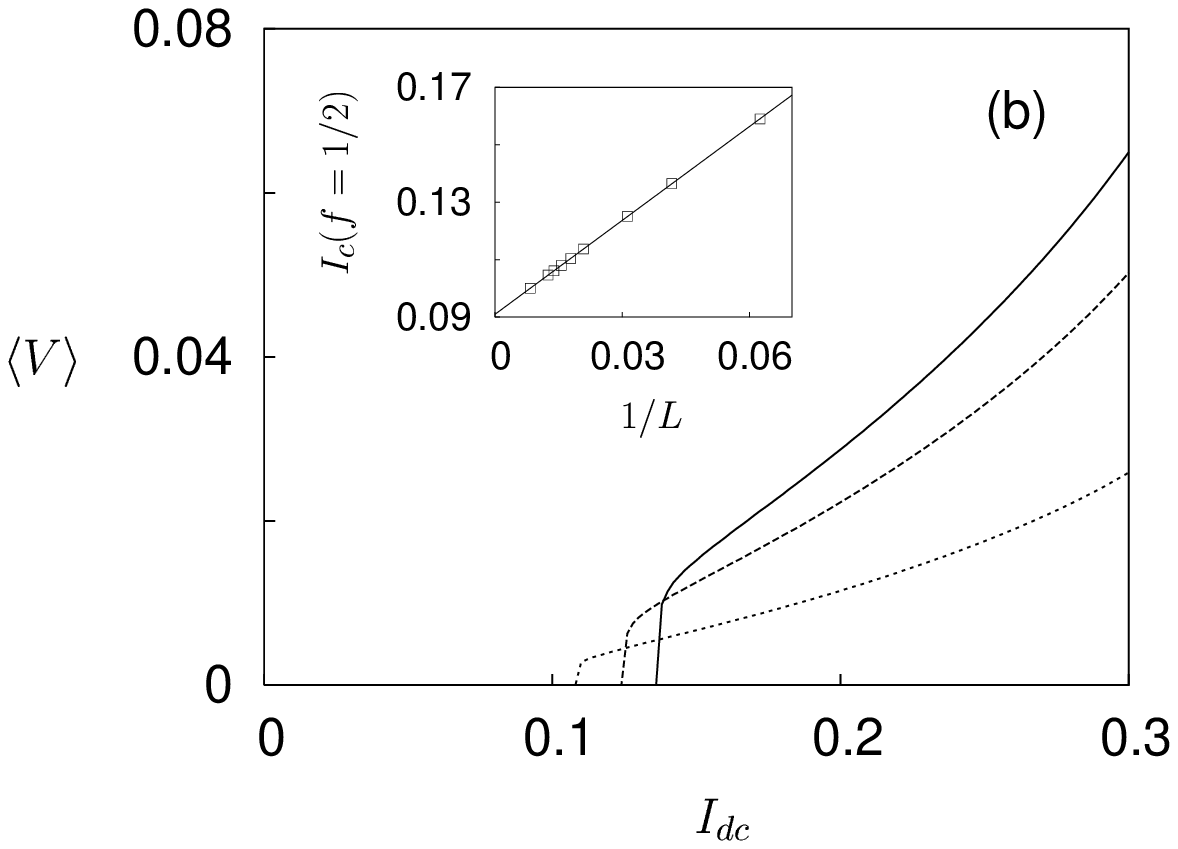}
\caption[$IV$ characteristics]
{$IV$ characteristics at (a) $f=0$ and (b) $f=1/2$ for the system size $L=$ 24, 32, and
64, respectively, from top to bottom. $I_{dc}$ and $\langle V\rangle$ are expressed
in units of $i_c$ and $i_c R$, respectively. Insets show critical currents
$I_c$ as functions of the system size $L$.} \label{fig:IV}
\end{figure}

Figure~\ref{fig:IV} shows the current-voltage $(IV)$ characteristics for (a) $f=0$
and (b) $f=1/2$ in the system of size $L = $ 24, 32, and 64 (from top to
bottom). It is observed that nonzero voltage drop develops as $I_{dc}$ is increased beyond 
the critical current $I_c$, the size dependence of which are shown in the insets. 
As $L \rightarrow \infty$, the critical currents for defect motion unambiguously approach
well-defined values from above\cite{footnote} and the
two-parameter least-square fits yield
\begin{eqnarray*}
I_c(f{=}0) & = &  0.101(1) \\
I_c(f{=}1/2) &=&  0.091(1),
\end{eqnarray*}
respectively, where the numbers in the parenthesis denote the
numerical errors in the last digits.
Our obtained value $I_c(0)$ is essentially the same as the known value,\cite{Rzchowski}
confirming the validity of our modified FTBC. 
For $f=1/2$, however, our result sharply disagrees with the numerical value
$I_c(1/2) \approx 0.32$ in Ref.~\onlinecite{Rzchowski} and appears 
much smaller than that of the defect-free system.\cite{bjkim:Ic} 
This indicates that the critical current to make the point defect mobile
is much less than the corresponding current for the rigid motion of the vortex 
superlattice. 
Remarkably, our results of the critical current are consistent with 
the experimental result $I_c \approx 0.1$, 
below which the dynamical resistance is negligible for any value of $f$.\cite{Benz} 
It is to be noted here that our results are obtained from the direct numerical calculation of
the $IV$ characteristics while other studies present mostly indirect estimates 
from the pinning energy barrier with the phase configuration of given vorticities 
assumed (see below).\cite{Rzchowski}

\begin{figure}
\centering \epsfig{width=0.4\textwidth,file=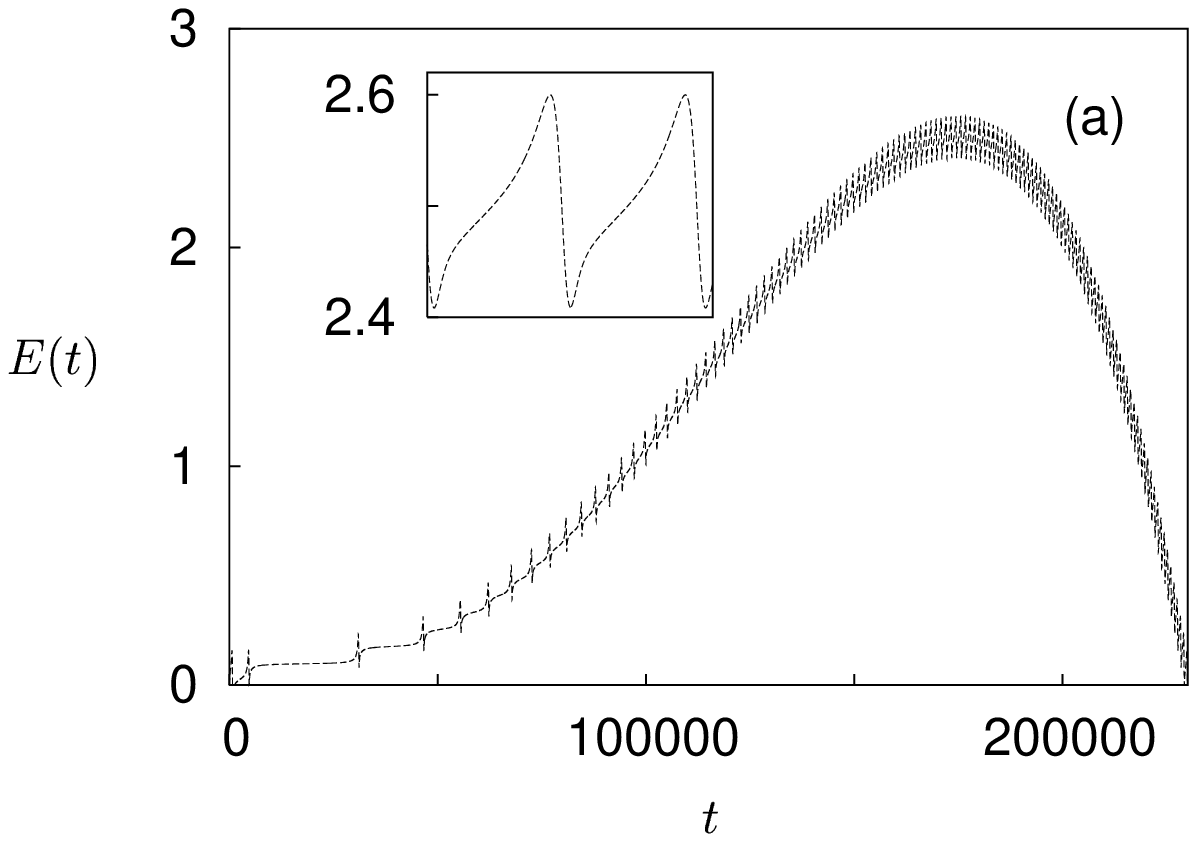}
\vspace{0.3cm} 
\epsfig{width=0.4\textwidth,file=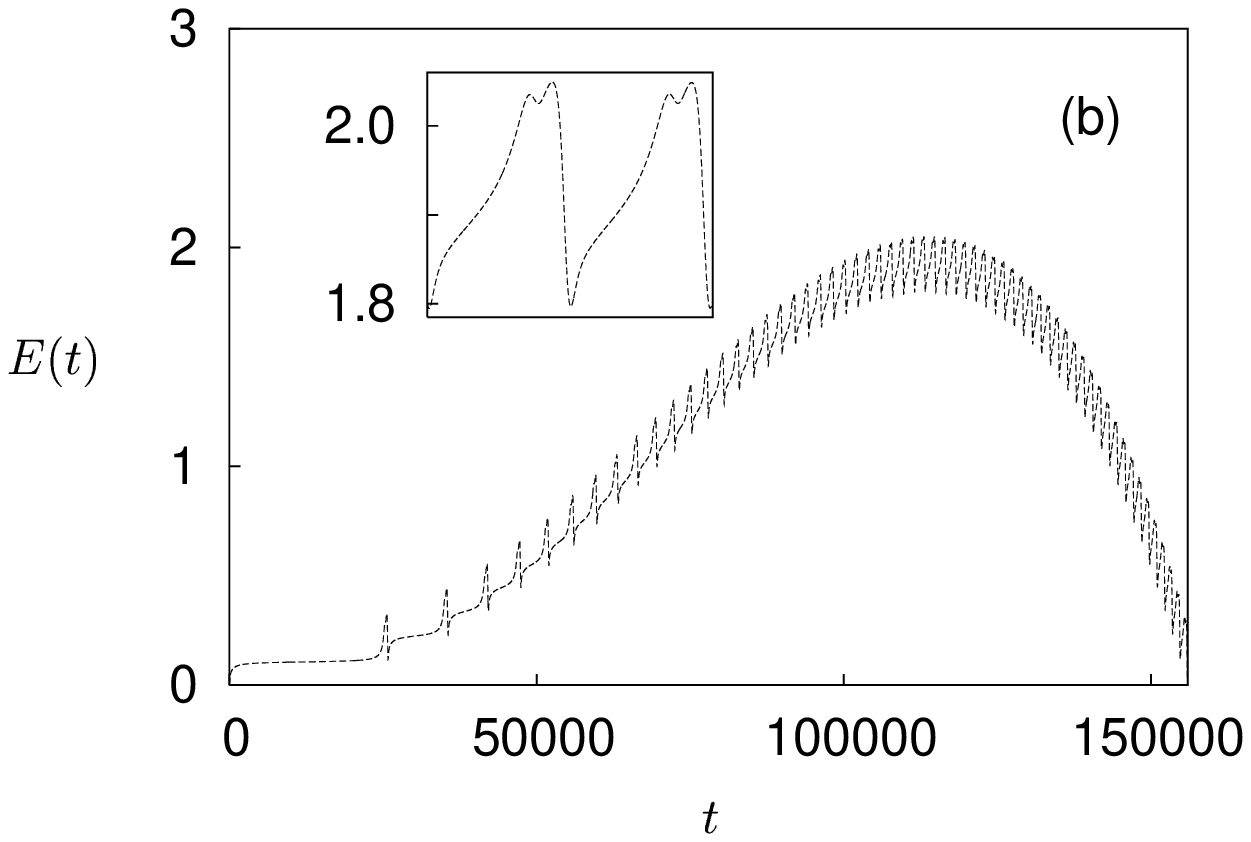}
\caption[Energy versus time]{Energy $E(t)$ versus time $t$ for
(a) $f=0$ and (b) $f=1/2$ in the system of size $L=120$ at injected currents 
$I_{dc} = 0.1128$ for $f=0$ and $0.1001$ for $f=1/2$. 
For convenience, $E(t)$ has been shifted such that $E=0$ is the minimum. 
Insets display expansion plots of $E(t)$ near maxima.}
\label{fig:Et}
\end{figure}

Our simulations also allow us to directly measure the pinning
energy barrier. To this end, we first plot in Fig.~\ref{fig:Et} the energy in Eq.~(\ref{eq:E})
as a function of time in the array of size $L=120$ at the external current 
slightly higher than the critical value: $I_{dc} = I_c(f,L) + 0.0001$. 
Observed are oscillatory behaviors of $E(t)$ in two very different time scales. 
While the long-period oscillations (only one period of which is shown for simplicity) 
describe the global motion of the defect,
the small oscillations shown in insets disclose the defect motion in short length scales.
The former arise from the defect motion across the system, i.e.,
the defect eventually arriving at one end of the system and then entering again 
from the other end. 
The lower energy values at both ends of the system reflect boundary effects, and the
asymmetry in the behavior manifests that the symmetry of the potential energy
around the middle of the system is broken by the driving currents.

On the other hand, the small oscillations in the insets are directly related with
the lattice pinning effects. 
The local minimum (dip) of the energy corresponds to the (extra) vortex sitting down 
at the center of a plaquette whereas the peak is produced by the vortex on a link in 
the course of moving on to the adjacent plaquette. 
In the case of $f=0$, there appears only one peak in a period. 
Since there are $L$ lattice pinning barriers for a defect to move across the system,
the longer oscillation in Fig.~\ref{fig:Et}(a) consists of $L$ small oscillations.  
For $f=1/2$, although overall features are similar to those for $f=0$, 
the shape of the pinning barrier is markedly different [see the inset of Fig.~\ref{fig:Et}(b)]: 
One small oscillation consists of double peaks separated by a small dip and 
these peaks are separated by bigger dips. The number of bigger dips turns out to be $L/2$, 
which reveals that the defect motion for $f=1/2$ is periodic in space with the period 
twice the lattice constant; this reflects the $2\times 2$ translational symmetry of the
vortex superlattice structure in the ground state. 
We define the pinning energy barrier $E_B$ to be the energy difference
between the bigger dip and the peak, neglecting the smaller one. 

\begin{figure}
\centering \epsfig{width=0.4\textwidth,file=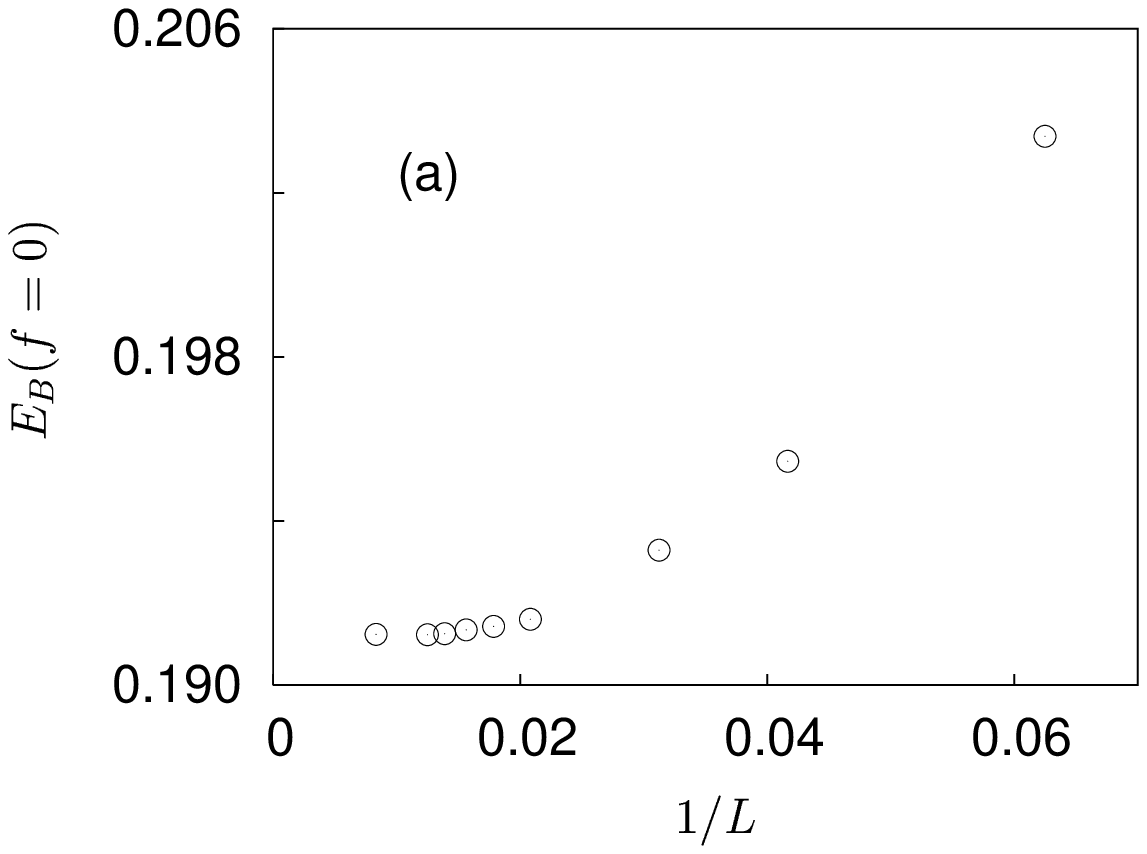}
\vspace{0.3cm} 
\epsfig{width=0.4\textwidth,file=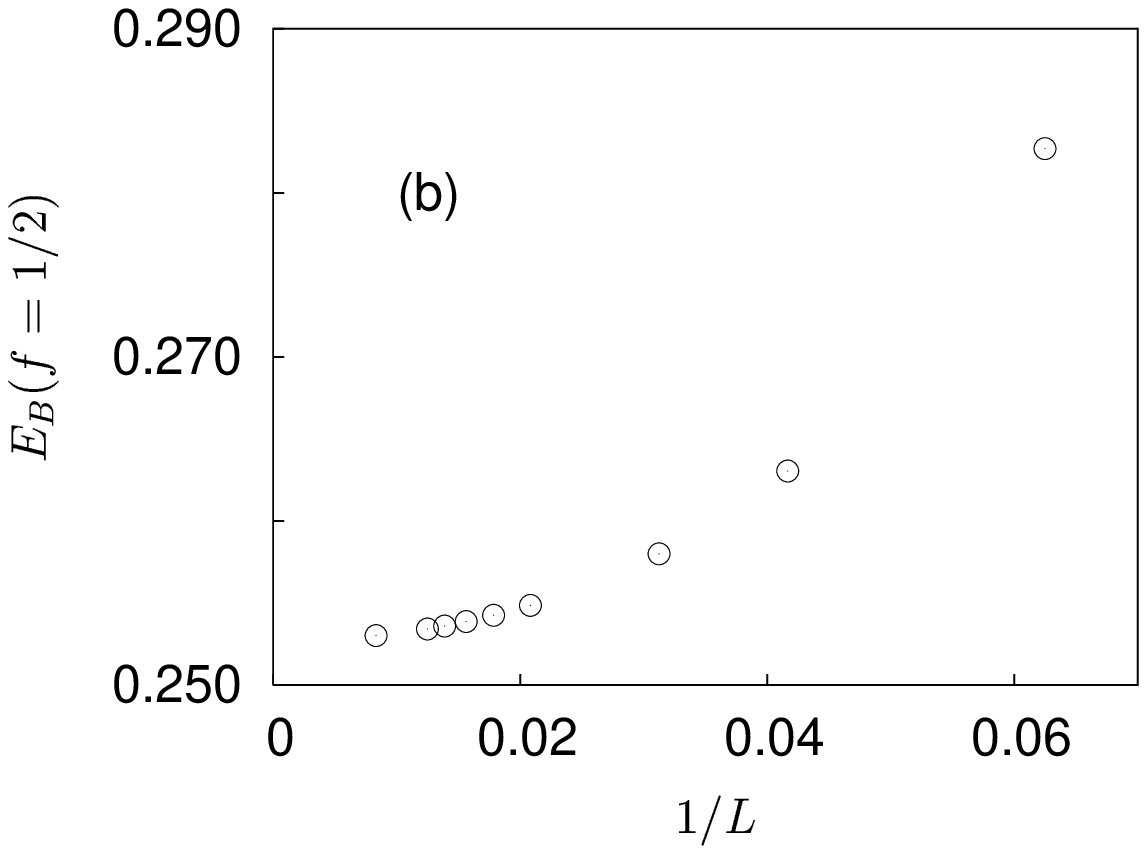}
\caption[Pinning barriers as a function of system
size]{Pinning barrier height $E_B(f)$ as a function of the system
size $L$ for (a) $f=0$ and (b) $f=1/2$. Systems of size up to
$L=120$ are displayed. } \label{fig:EL}
\end{figure}

To minimize the boundary effects, we measure the barrier at the maximum of the long-period
oscillation, namely when the defect resides near the center of the system.
Figure~\ref{fig:EL} presents the measured barrier height versus the system size. 
The extrapolation to the thermodynamic limit leads to~\cite{com}
\begin{eqnarray*}
E_B(f{=}0) &=& 0.192(1) \\
E_B(f{=}1/2) &=& 0.255(1).
\end{eqnarray*}
The resulting ratio, $E_B(1/2)/E_B(0) \approx 1.33$, is again
in good agreement with the experimental result in Ref.~\onlinecite{Rzchowski}. 
Note the dramatic improvement over the existing value, 
$E_B(1/2)/E_B(0) \approx 6$, obtained via the iteration method.
%

\begin{figure}
\centering \epsfig{width=0.45\textwidth,file=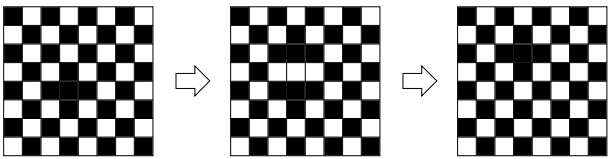}
\caption[Pattern of vortices motion]
{Pattern of the defect motion in the system with $f=1/2$. Currents are applied along
the $-x$ direction (i.e., from right to left). } \label{fig:motion}
\end{figure}

We now provide a physical explanation as to the origin of the previous discrepancy, 
in relation with the vortex motion directly observed in our simulations.  
We display in Fig.~\ref{fig:motion} the vortex motion in the $f=1/2$ background, 
which reveals the two-step process in the defect motion: 
The defect first pushes away the nearby vortex to make the plaquette vacant and 
subsequently occupies this vacancy. This type of motion gives rise to small oscillations 
of the energy in Fig.~\ref{fig:Et}.
In the numerical estimate in Ref.~\onlinecite{Rzchowski}, however, 
the vortex defect was assumed to sit on the top of the other (background) vortex,
which results in much higher energy due to strong on-site repulsion. 
This type of vortex configuration implies that the extra vortex moves in the rigid
vortex lattice background, which is different from the actual motion of the vortices 
shown in Fig.~\ref{fig:motion}. We believe that the latter type of actual motion occurs
in experiment. Namely, vortex motion is influenced by background vortices
present in the frustrated system, thus yielding substantially lower critical currents. 
%
%

In summary, we have introduced novel boundary conditions which are convenient 
to study vortex dynamics in JJAs. From dynamic simulations in the presence of external 
currents and magnetic fields, the critical currents as well as the pinning energy
barriers to vortex motion have been directly measured without any a priori assumption 
on the vortex configuration. 
For $f=0$, our results agree fully with previous theoretical and experimental results. 
For $f=1/2$, our results also give excellent agreement with experimental ones, in contrast
with the previous theoretical prediction. 
>From the direct observation of the vortex motion, it has been found that for $f=1/2$ the
defect moves in the two-step process with two vortices involved, 
producing double peaks in the energy barrier. This suggests that vortices move 
interactively when a large number of vortices are present.

This work was supported in part by the KOSEF Grant R01-2002-000-00285
and the SKOREA program (JSL, MYC) and by the 
KRF Grant 2003-041-C00137 (BJK).

\end{document}